\documentclass[a4paper]{jpconf}
\usepackage{graphicx}

\usepackage{amsmath}
\usepackage{amssymb}
\usepackage{amsthm}

\newcommand{\pol}{\varepsilon}

\newcommand{\cZ}{{\cal Z}}

\def\sdel{{\underset{\smile}{\Delta}}}
\def\bcdel{{}^\circ\! {\underset{\smile}{\Delta}}}
\def\bddel{{}^\bullet\! {\underset{\smile}{\Delta}}}

\def\bddeld{{}^{\bullet}\! {\underset{\smile}{\Delta}^{\hskip -.5mm \bullet}}}

\def\bcdeld{{}^{\circ}\! {\underset{\smile}{\Delta}^{\hskip -.5mm \bullet}}}

\def\half{{1\over 2}}

\def\fourth{{1\over4}}

\def\Z{{\mathchoice {\hbox{$\sf\textstyle Z\kern-0.4em Z$}}
{\hbox{$\sf\textstyle Z\kern-0.4em Z$}}
{\hbox{$\sf\scriptstyle Z\kern-0.3em Z$}}
{\hbox{$\sf\scriptscriptstyle Z\kern-0.2em Z$}}}}

\def\square{\kern1pt\vbox{\hrule height 1.2pt\hbox{\vrule width 1.2pt
   \hskip 3pt\vbox{\vskip 6pt}\hskip 3pt\vrule width 0.6pt}
   \hrule height 0.6pt}\kern1pt}
      \def\boxop{{\raise-.25ex\hbox{\square}}}

\def\mn{{\mu\nu}}

\def\Tr{{\rm Tr}\,}

\def\e{\,{\rm e}}

\def\be{\begin{equation}}
\def\ee{\end{equation}\noindent}
\def\bear{\begin{eqnarray}}
\def\ear{\end{eqnarray}\noindent}
\def\bec{\blue\begin{equation}}
\def\eec{\end{equation}\black\noindent}
\def\bearc{\blue\begin{eqnarray}}
\def\earc{\end{eqnarray}\black\noindent}
\def\benn{\begin{enumerate}}
\def\enn{\end{enumerate}}

\def\totint{\int_{-\infty}^{\infty}}

\def\slash#1{#1\!\!\!\raise.15ex\hbox {/}}
\newcommand{\slD}{\,\raise.15ex\hbox{$/$}\kern-.27em\hbox{$\!\!\!D$}}
\newcommand{\slpartial}{\raise.15ex\hbox{$/$}\kern-.57em\hbox{$\partial$}}

\newcommand{\nc}{\newcommand}
\nc{\spa}[3]{\left\langle#1\,#3\right\rangle}
\nc{\spb}[3]{\left[#1\,#3\right]}
\nc{\ksl}{\not{\hbox{\kern-2.3pt $k$}}}
\nc{\hf}{\textstyle{1\over2}}
\nc{\tq}{{\tilde q}}
\nc{\esl}{\not{\hbox{\kern-2.3pt $\pol$}}}

\def\kinb{{1\over 4}\dot x^2}

\def\4piTD{{(4\pi T)}^{-{D\over 2}}}
\def\4piT4{{(4\pi T)}^{-2}}

\def\Tintm4{{\dps\int_{0}^{\infty}}{dT\over T}\,e^{-m^2T}
    {(4\pi T)}^{-2}}
\def\Tintm{{\dps\int_{0}^{\infty}}{dT\over T}\,e^{-m^2T}}

\newcommand{\slG}{{{\dot G}\!\!\!\! \raise.15ex\hbox {/}}}

\def\GBd12{{\dot G}_{B12}}

\newcommand{\no}{\noindent}

\def\dps{\displaystyle}

\begin{document}
\title{
The worldline formalism in strong-field QED
}

\author{Christian Schubert}

\address{Centro Internacional de Ciencias, A.C. Campus UNAM-UAEM, Avenida Universidad 1001,
62100 Cuernavaca, Morelos, Mexico. 

After Sepember 15, 2022: ELI Beamlines Centre, 
Institute of Physics, Czech Academy of Sciences,
Za Radnicki 835, 25241 Dolni B\v{r}e\v{zany}, Czech Republic.
}

\ead{christianschubert137@gmail.com}

\begin{abstract}
The worldline formalism provides an alternative to Feynman diagrams that has been found particularly useful for external-field calculations in quantum electrodynamics.
Here I summarize its present range of applications, which
includes Schwinger pair creation, photon splitting in constant fields and plane-wave backgrounds, as well
as non-linear Compton scattering in constant fields.
\end{abstract}

\vspace{-20pt}

\section{The trouble with strong-field QED}

Since strong background fields must be treated non-perturbatively, QED calculations in the presence of such fields 
are usually done using Feynman rules involving the exact electron propagator in the field, and on-shell polarization eigenstates in the field.
Of course, a closed-form expression for this propagator is available only for a limited number of field configurations. 
And even in these fortuitous cases, in practice those generalized Feynman rules lead to exceedingly cumbersome calculations. 
In the words of the authors of a well-known textbook \cite{ditgie-book},
“it appears doubtful to us that the formulation of spinor electrodynamics employing the usual Dirac algebra is appropriate for higher-order external-field calculations. 
Even computer-aided algebraic-manipulation programs seem to be of limited use”. 

Presently, the main alternative to the Feynman diagram approach is the ``worldline formalism''. In this talk, I give a short review of its present state of development
for QED calculations involving constant and plane-wave fields, as well as for Schwinger pair creation in general electric fields. However, let us start with some history. 
 
\section{The worldline formalism in QED}

The worldline formalism was invented by Feynman simultaneously with Feynman diagrams \cite{feynman1950,feynman1951}, 
but seriously developed as a competitor to the diagrammatic approach only during the nineties
in the context of QCD amplitudes and their representation as the field-theory limit of string amplitudes \cite{polyakov-book,berkosPRL,berkosNPB,strassler1}. 
It has  turned out to be particularly well-suited to the study of QED processes in external fields \cite{5,shaisultanov,17,18,141}. 

\begin{figure}[htbp]
\begin{center}
 \includegraphics[width=0.45\textwidth]{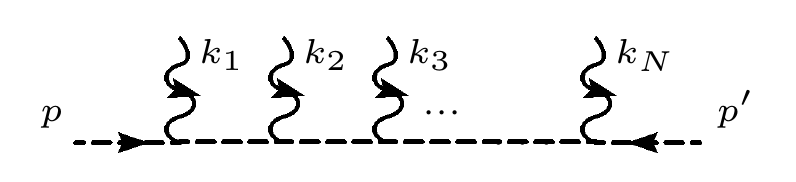}
\caption{{Photon-dressed scalar propagator in momentum space. }}
\label{fig-propexpand}
\end{center}
\end{figure}

As far as QED is concerned, the starting point is the following path-integral representation of the $x$-space propagator of a relativistic charged scalar particle in
an external field $A^\mu(x)$,  
\bear
D^{xx'} [A] &=& 
\langle x'| \int_0^{\infty}dT\, {\rm exp}\Bigl\lbrack - T (  -(\partial + ie A)^2+m^2)\Bigr\rbrack  | x \rangle
\nonumber\\
&=&
\int_0^{\infty}
dT\,
\e^{-m^2T}
\int_{x(0)=x}^{x(T)=x'}
{\cal D}x(\tau)\,
e^{-\int_0^T d\tau \bigl(\kinb +ie\dot x\cdot A(x(\tau))\bigr)}
\label{scalpropfreepi}
\ear\no
Here the path integral is over all trajectories running in the fixed proper-time $T$ from $x$ to $x'$ in (euclidean) spacetime.
Expanding the field in $ N$ plane waves,
\bear
A^{\mu}(x(\tau)) = \sum_{i=1}^N \,\varepsilon_i^{\mu}\e^{ik_i\cdot x(\tau)}
\label{pw}
\ear
and Fourier transforming the endpoints one gets a path-integral representation of the ``$N$-photon-dressed propagator''
in momentum space, Fig. \ref{fig-propexpand},  where  summation over all permutations of the photons is understood.

Similarly, one has the following representation for the one-loop effective action in scalar QED:
\bear
\Gamma [A] &=&- {\rm \Tr}{\rm ln} \Bigl\lbrack -(\partial + ie A)^2+m^2\Bigr\rbrack  
= \int_0^{\infty} \frac{dT}{T} \,{\rm Tr} \, {\rm exp}\Bigl\lbrack - T (  -(\partial + ie A)^2+m^2)\Bigr\rbrack  \nonumber\\
&=&
\int_0^{\infty}
\frac{dT}{T}\,
\e^{-m^2T}
\int_{x(0)=x(T)}
{\cal D}x(\tau)\,
e^{-\int_0^T d\tau \bigl(\kinb +ie\dot x\cdot A(x(\tau))\bigr)}
\label{EA}
\ear
The path integral is now over all closed loops with periodicity $T$ in the proper time of the scalar in the loop. 
The plane-wave ansatz (\ref{pw}) then
leads to a path-integral representation of the one-loop $N$ - photon amplitudes, again already summed over all permutations.

In \cite{feynman1951}, Feynman found a generalization of (\ref{EA}) to the spinor QED case where the electron spin is
implemented simply by adding, under the path integral, the ``spin factor'' ${\rm Spin}[x,A]$,
\bear
{\rm Spin}[x,A] = {\rm tr}_{\Gamma} {\cal P}
\exp\Biggl[{{i\over 4}e\,[\gamma^{\mu},\gamma^{\nu}]
\int_0^Td\tau F_{\mu\nu}(x(\tau))}\Biggr]
\ear
Here $\cal P$ denotes path ordering which will now have to be imposed for generic $F_\mn$. To a large extent, this spoils
the elegance as well as practical usefulness of the formalism (and may have been a reason for Feynman not to pursue
further this formalism). Following the advent of Grassmann variables in the sixties, Fradkin \cite{fradkin} showed that 
it can be avoided replacing the Feynman spin factor by the following path integral,
\bear
{\rm Spin}[x,A]  =
\int {\cal D}\psi(\tau)
\,
\exp 
\Biggl\lbrack
-\int_0^Td\tau
\Biggl(
\half \psi\cdot \dot\psi -ie \psi^{\mu}F_{\mn}\psi^{\nu}
\Biggr)
\Biggr\rbrack
\ear
where $\psi^\mu(\tau)$ is a Lorentz-vector valued function of proper time with components that are anticommuting functions,
$\psi^\mu(\tau_1)\psi^\nu(\tau_2) = - \psi^\nu(\tau_2)\psi^\mu(\tau_1)$, and the Grassmann path integral has to be evaluated with anti-periodic boundary conditions,
$\psi^\mu(T) = - \psi^\mu(0)$.

While for the closed-loop case the choice of the spin-factor representation is essentially the only decision to be made, for the open fermion line
there are further choices to be made originating from the boundary conditions for the spin degrees of freedom \cite{feynman1951,fradkin,fragit,18}.
Quite recently, a worldline representation of the $x$-space Dirac propagator in a Maxwell field that seems promising for many applications has been constructed 
in \cite{130,131}
\bear
S^{xx'}[A] &=&
\bigl[m + i\slash{D}'\bigr]
K^{xx'}[A]\,\nonumber\\
K^{xx'}[A] &=&
\int_0^{\infty}
dT\,
\e^{-m^2T}
\e^{-\fourth \frac{(x-x')^2}{T}}
\int_{q(0)=0}^{q(T)=0}
Dq\,
\e^{-\int_0^T d\tau\bigl(
\fourth \frac{\dot q^2}{4}
+ie\,\dot q\cdot A 
+ie \frac{x'-x}{T}\cdot A
\bigr)}
\nonumber\\
&& \times\,  2^{-\frac{D}{2}}
{\rm symb}^{-1}
\int_{\psi(0)+\psi(T)=0
\hspace{-30pt}}D\psi
\, \e^
{-\int_0^Td\tau\,
\bigl[\half\psi_{\mu}\dot\psi^{\mu}-ieF_{\mu\nu}(\psi+\eta)^{\mu}(\psi+\eta)^{\nu}\bigl]
}
\ear\no
Here $\eta^{\mu}$ is an external anticommuting Lorentz vector, and the ``symbol map''  {\it symb}  converts products of $\eta$'s 
into antisymmetrised products of Dirac matrices:

\bear
{\rm symb} 
\bigl(\gamma^{\alpha_1\alpha_2\cdots\alpha_n}\bigr) \equiv 
(-i\sqrt{2})^n
\eta^{\alpha_1}\eta^{\alpha_2}\ldots\eta^{\alpha_n}
\ear
%

where $\gamma^{\alpha\beta\cdots\rho}$ denotes the totally antisymmetrised product of gamma matrices:

\bear
\gamma^{\alpha_1\alpha_2\cdots \alpha_n} \equiv \frac{1}{n!}\sum_{\pi\in S_n} 
{\rm sign}(\pi) \gamma^{\alpha_{\pi(1)}}\gamma^{\alpha_{\pi(2)}} \cdots \gamma^{\alpha_{\pi(n)}} 
\ear

\section{String-inspired treatment of the worldline path integral}

The modern approach to the worldline formalism came up in the nineties when, following the development of string perturbation
theory, worldline path integrals started to be seen as one-dimensional analogs of Polyakov's worldsheet path integral (see
\cite{berntasi,41} for more on the history of the subject). Thus in the ``string-inspired'' version of the worldline formalism, all path integrals
are, be it by approximation, expansion or transformation, manipulated into gaussian ones, which reduces their evaluation to the
determination of appropriate worldline Green's functions and the standard combinatorics of Wick contractions. 
 In the QED case, the appropriate worldline correlator for the worldline coordinate and Grassmann fields are 
\bear
\langle x^{\mu}(\tau_1)x^{\nu}(\tau_2) \rangle
&=&
-G(\tau_1,\tau_2)\, \delta^{\mu\nu} \, , \quad
G(\tau_1,\tau_2) = \vert \tau_1 -\tau_2\vert - \frac{\Bigl(\tau_1 -\tau_2\Bigr)^2}{T} 
\\
\langle \psi^{\mu}(\tau_1)\psi^{\nu}(\tau_2)\rangle
&=&
\half
G_F(\tau_1,\tau_2)\, \delta^{\mu\nu} \, , \quad
G_F(\tau_1,\tau_2) = {\rm sign}(\tau_1 - \tau_2)
\label{wickferm}
\ear
This allows one to write down compact Koba-Nielsen type master formulas for the one-loop $N$-photon amplitudes
in scalar and spinor QED \cite{polyakov-book, berkosPRL,berkosNPB,strassler1} and (with a different Green's function for the coordinate path integral) 
also for the corresponding photon-dressed propagators \cite{dashsu,130,131}. 
The simplest one of these various master formulas is the one for the $N$-photon amplitude in scalar QED,
\begin{eqnarray}
\Gamma[\lbrace k_i,\varepsilon_i\rbrace]
&=&
{(-ie)}^N
{\dps\int_{0}^{\infty}}{dT\over T}
{(4\pi T)}^{-{D\over 2}}
e^{-m^2T}
\prod_{i=1}^N \int_0^T 
d\tau_i
\nonumber\\
&&\hspace{-8pt}
\times
\exp\biggl\lbrace\sum_{i,j=1}^N 
\bigl\lbrack \half G_{ij} k_i\cdot k_j
+i\dot G_{ij}k_i\cdot\varepsilon_j 
+\half\ddot G_{ij}\varepsilon_i\cdot\varepsilon_j
\bigr\rbrack\biggr\rbrace
\mid_{{\rm lin}(\varepsilon_1,\ldots,\varepsilon_N)}
\nonumber\\
\label{master}
\end{eqnarray}
\no
Here $\tau_i$ parametrizes the position of photon $i$ along the loop,
and we have abbreviated $G_{ij} \equiv G(\tau_i,\tau_j)$, $\dot G_{ij}\equiv \frac{\partial}{\partial \tau_i} G_{ij}$, etc. 
The notation $\mid_{{\rm lin}(\varepsilon_1,\ldots,\varepsilon_N)}$ means that only terms linear in each of the photon 
polarization vectors are to be kept. 
This master formula holds the full information on the $N$-photon amplitudes in scalar QED (on-shell and off-shell). 

A formally analogous master formula can be written down for the spinor loop using a one-dimensional superspace formalism,
but in practice it is usually preferable to use a pattern-matching procedure, discovered (in the QCD context) by Bern and Kosower \cite{berkosPRL,berkosNPB}.
Namely, it is possible to remove all the second derivatives of the Green's function $G$ through integration-by-parts (this can be done
maintaining the full permutation symmetry between the $N$ photons \cite{26}), after which, up to the global normalization, all the effects of the electron spin
can be taken into account by applying, simultaneously to all ``$\tau$-cycles'' 
$ \dot G_{i_1i_2}  \dot G_{i_2i_3}  \cdots \dot G_{i_ni_1}$ in the resulting integrand, the replacement rule
\bear
\dot G_{i_1i_2} 
\dot G_{i_2i_3} 
\cdots
\dot G_{i_ni_1}
\rightarrow 
\dot G_{i_1i_2} 
\dot G_{i_2i_3} 
\cdots
\dot G_{i_ni_1}
-
G_{Fi_1i_2}
G_{Fi_2i_3}
\cdots
G_{Fi_ni_1}
\label{rr}
\ear
In the original string-based formalism of \cite{berkosPRL,berkosNPB} this rule comes from worldsheet supersymmetry.
In the worldline formalism it relates to a formal supersymmetry between the electron spin and orbital degrees of freedom,
familiar from the non-relativistic Pauli equation (see, e.g., \cite{cksbook}). 

\section{QED in a constant field}

Let us now come to the main topic of this talk, QED in external fields, starting with the constant-field case. 
Quite analogously to the Feynman-diagrammatic approach, where the effect of a constant background field
can be absorbed through a change of the electron propagator, in the worldline formalism it can be taken
care of by the following changes of the worldline Green's functions and the global path-integral normalization factors
\cite{5,shaisultanov,18,40}:

\benn

\item
Change the worldline Green's functions $ G, G_F$ to field-dependent ones $ {\cal G}_B, {\cal G}_F$, 

\begin{eqnarray}
G(\tau_1,\tau_2) &\to &
{\cal G}_{B}(\tau_1,\tau_2) = \frac{T}{2{\cal Z}^2}
\biggl({{\cal Z}\over{{\rm sin}{\cal Z}}}
\,{\rm e}^{-i{\cal Z}\dot G_{12}}
+i{\cal Z}\dot G_{12} -1\biggr)
\nonumber\\
G_F(\tau_1,\tau_2) &\to &
{\cal G}_{F}(\tau_1,\tau_2) =
G_{F12}
{{\rm e}^{-i{\cal Z}\dot G_{12}}\over {\rm cos}{\cal Z}}
\nonumber\\
\label{change}
\end{eqnarray}
\noindent
where the right-hand sides are to be understood as power series in the Lorentz matrix $ {\cal Z}_{\mu\nu} \equiv eF_{\mu\nu}T$ (see \cite{40,41}
for more practically useful representations).

\item
Add factors ${\rm det}^{{1\over 2}}\bigl\lbrack\frac{\cal Z}{{\rm sin}{\cal Z}}\bigr\rbrack$ for the coordinate path integral
and ${\rm det}^{{1\over 2}}\lbrack {\rm cos}{\cal Z}\rbrack $ for the Grassmann path integral. 

\enn

For example, applying these rules to the master formula (\ref{master}) transforms it into the following master formula for the
scalar QED $N$-photon amplitudes in a constant field \cite{shaisultanov,18}:
\begin{eqnarray}
&&\Gamma_{\rm scal}
[k_1,\varepsilon_1;\ldots;k_N,\varepsilon_N;F]
=
{(-ie)}^N
{\dps\int_{0}^{\infty}}{dT\over T}
{(4\pi T)}^{-{D\over 2}}
e^{-m^2T}
{\rm det}^{{1\over 2}}
\biggl[
\frac{\cal Z}{{\rm sin}{\cal Z}}
\biggr]
\prod_{i=1}^N \int_0^T 
d\tau_i
\nonumber\\
&&
\times
\exp\biggl\lbrace\sum_{i,j=1}^N 
\Bigl\lbrack \half k_i\cdot {\cal G}_{Bij}\cdot  k_j
-i\varepsilon_i\cdot\dot{\cal G}_{Bij}\cdot k_j
+\half
\varepsilon_i\cdot\ddot {\cal G}_{Bij}\cdot\varepsilon_j
\Bigr\rbrack\biggr\rbrace
\mid_{{\rm lin}(\varepsilon_1,\ldots,\varepsilon_N)}
\label{masterF}
\end{eqnarray}
\no
For the transition to the spinor-loop case amplitudes one still has the same choice as in the vacuum case, i.e. between the introduction of
a worldline super formalism or an appropriate generalization of the replacement rule (\ref{rr}) \cite{41}. 

The determinant factors ${\rm det}^{{1\over 2}}\bigl\lbrack\frac{\cal Z}{{\rm sin}{\cal Z}}\bigr\rbrack$ and ${\rm det}^{{1\over 2}}\bigl\lbrack\frac{\cal Z}{{\rm tan}{\cal Z}}\bigr\rbrack$
by themselves just produce the vacuum amplitudes in the constant field, i.e. the Weisskopf \cite{weisskopf}  and Euler-Heisenberg \cite{eulhei} Lagrangians. 

The constant-field master formula (\ref{masterF}) was then applied to recalculations of the vacuum polarization tensors in a general constant field \cite{ditsha,40} 
and of the photon-splitting amplitudes in a magnetic field \cite{17}, for both scalar and spinor QED. More recently, it was also used for the calculation
of the one-photon (tadpole) amplitudes \cite{112,113}, which are irrelevant at the one-loop level but have attracted attention following the recent demonstration by
Gies and Karbstein that they cannot be neglected when appearing as subdiagrams at the multiloop level \cite{giekar}.

\begin{figure}[htbp]
\begin{center}
 \includegraphics[width=0.35\textwidth]{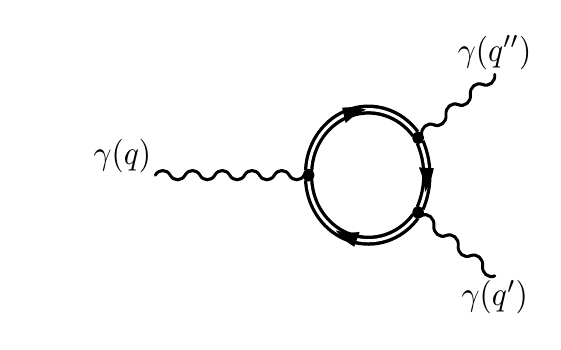}
\caption{{Photon splitting in a magnetic field. }}
\label{fig-phosplit}
\end{center}
\end{figure}

The magnetic photon-splitting amplitude (Fig.  \ref{fig-phosplit}), calculated in the standard approach in 1971 by Adler \cite{adler71} and recalculated in 1996 by Adler and the author
in the worldline formalism \cite{17,adler-book} provides a good example for pinpointing the main advantages of the latter approach:

\benn

\item
The loop momentum has already been integrated out from the beginning at the level of the construction of the coordinate path integral.
 
\item
The loop is parametrized globally from the beginning, rather than segmented into three propagators with their individual proper-time
parametrizations. 

\item
This also has made it possible to avoid fixing an ordering for the three photons along the loop.

\item
Specifically for scalar QED: the same integration-by-parts that has led to the replacement rule (\ref{rr}) has effectively
also removed the quartic seagull vertex, which is a nuisance in the diagrammatic approach. 

\item
Specifically for spinor QED: the considerable algebraic work caused by the nontrivial Dirac matrix structure of the generalized electron
propagator has been replaced by the simple pattern-matching procedure (\ref{rr}) (now involving the generalized Green's functions (\ref{change})). 

\enn

All these advantages will gain significance with increasing $N$. 

The final three-parameter integrals are of the same type as
the ones found in the standard approach, but far from identical. They have to be done numerically.

In \cite{18}, the master formula (\ref{masterF}) was generalized to the multiloop level and applied to a calculation
of the two-loop Euler-Heisenberg Lagrangian. For the case of a self-dual background field, this eventually led
to a closed-form result \cite{50}.

For a generalization to Einstein-Maxwell theory see \cite{61,71,144}, where the worldline formalism has ben applied
to the calculation of the one-loop photon-graviton amplitude with a scalar and spinor loop. 

\section{Photon-dressed scalar propagator in a constant field}

Generalization to the open scalar line, i.e. the derivation of a master formula for the  $N$-photon dressed scalar propagator in a constant
field, has been achieved in \cite{110}. In momentum space, this formula reads
\bear
D^{pp'}(k_1,\varepsilon_1;\cdots;k_N,\varepsilon_N;F) &=& 
(-ie)^N (2\pi)^{D} \delta \Bigl( p+ p^\prime+ \sum^{N}_{i=1}k_{i}\Bigr)   \int^{\infty}_{0}dT \e^{-m^2T}
\frac{1}{{\det}^{\half}\left[{\rm cos\cZ} \right]}
\nonumber\\ 
&&\hspace{-90pt}\times\int^{T}_{0}d\tau_{1}\cdots \int^{T}_{0}d\tau_{N}
\, \e^{\sum^{N}_{i,j=1}\bigl( k_{i}\cdot \sdel{}_{ij}\cdot k_{j}
-2i\varepsilon_{i}\cdot \bddel{}_{ij}\cdot k_{j}
-\varepsilon_{i}\cdot \bddeld{}_{ij}\cdot \varepsilon_{j}\bigr)} 
\e^{-T b \cdot (\frac{{\rm tan\cZ}}{\cZ}) \cdot b}
\Big\vert_{\varepsilon_1\varepsilon_2\cdots \varepsilon_N}
\nonumber\\
\label{master-pspace}
\ear
The essential new ingredient here is the open-line worldline Green's function $\sdel(\tau,\tau^{\prime})$,
which is related to the closed-loop one $ {\cal{G}}_{B}(\tau,\tau^{\prime})$ by
\begin{eqnarray}
\sdel(\tau,\tau^{\prime})
=\half \Bigl( {\cal{G}}_{B}(\tau,\tau^{\prime})-{\cal{G}}_{B}(\tau, 0)-{\cal{G}}_{B}(0,\tau^{\prime})+{\cal{G}}_{B}(0,0)\Bigr)
\label{appB-defDeltafriendly} 
\end{eqnarray}
Contrary to the latter, it is not translation invariant, so that one has to distinguish between a derivative with respect to the
first (left `bullet') and second variable (right `bullet'). Similarly, a left (right) `open circle' on $\sdel(\tau,\tau')$ denotes an integral $\int_0^T d\tau$ ($\int_0^Td\tau'$). 
The Lorentz vector $b^\mu$ is defined by
\begin{eqnarray}
b \equiv p+\frac{1}{T}\sum^{N}_{i=1}\Bigl[\Bigl( 
  \tau_{i}-2ie F\cdot \bcdel{}_{i}\cdot \Bigr) k_i -i \left(1-2ie F\cdot \bcdeld{}_{i} \cdot \Bigr)  \varepsilon_{i} \right] 
  \label{defb}
\end{eqnarray}

The amplitude given by the master formula \eqref{master-pspace} corresponds to the set of Feynman diagrams depicted in Fig.~\ref{fig-propexpand},
only that now all the scalar propagators are the exact ones in the external field (usually indicated by a double line). 
When applying it to the calculation of physical processes, one has to take into account that it describes the {\it untruncated} dressed propagator,
i.e. the final propagators on each end of the scalar line in Fig.~\ref{fig-propexpand} are included. 
The $N=1$ case was applied in \cite{112} to the calculation of the ``Gies-Karbstein addendum'' for the one-loop scalar propagator,
the $N=2$ one in \cite{110} to obtain a compact integral representation for the Compton scattering cross section in a constant field in scalar QED.

For the further generalization of the master formula (\ref{master-pspace}) to the open fermion line, using the symbol formalism introduced in section 2, 
see \cite{113,ahedil,inprep}.

\section{QED in a plane-wave background}

In strong-field QED, equally important as the constant-field background is the plane-wave one, defined by a vector potential $A(x)$ of the form
\bear
e A_{\mu}(x) = a_{\mu}(n\cdot x)
\label{defpw}
\ear
where $n^{\mu}$ is a null vector, $ n^2 = 0$, and as is usual we will further impose the {\it light-front gauge condition} 
$
n\cdot a = 0
$.
Like the constant-field case it leads to Dirac and Klein-Gordon equations that can be solved exactly.
Nevertheless, it was treated in the worldline formalism only relatively recently, mainly because it was
not understood how to manipulate the corresponding worldline path integrals into gaussian form. 
Building on work on the vacuum polarization case by Ilderton and Torgrimsson \cite{ildtor}, 
the following master formula for the scalar QED $ N$-photon amplitude in a plane-wave background
was obtained in 2019 by J.P. Edwards and the author \cite{141}:
\bear
\Gamma_{\rm scal}(\lbrace{k_i,\varepsilon_i\rbrace};a) &=&
(-ie)^N 
(2\pi)^3 
\delta\bigl(\sum_{i=1}^N k_i^1\bigr)
\delta\bigl(\sum_{i=1}^N k_i^2\bigr)
\delta\bigl(\sum_{i=1}^N k_i^+\bigr)
\totint dx_0^+ \e^{-i x_0^+ \sum_{i=1}^N k_i^-}
\nonumber\\&&\hspace{-80pt}\times
\int_0^{\infty}
\frac{dT}{T}\,
{(4\pi T)}^{-{D\over 2}}
\prod_{i=1}^N \int_0^Td\tau_i
\e^{
\sum_{i,j=1}^N 
\bigl\lbrack  \half G_{ij} k_i\cdot k_j
-i\dot G_{ij}\varepsilon_i\cdot k_j
+\half\ddot G_{ij}\varepsilon_i\cdot\varepsilon_j
\bigr\rbrack}
\nonumber\\&&\hspace{-80pt}\times
\e^{- M^2 T +2\sum_{i=1}^N k_i \cdot 
\bigl(I(\tau_i)-\langle\langle I \rangle\rangle \bigr)
-2i \sum_{i=1}^N\bigl(a(\tau_i)-\langle\langle a\rangle\rangle \bigr) \cdot \varepsilon_i
}
\mid_{{\rm lin}(\varepsilon_1,\ldots,\varepsilon_N)}
\nonumber\\
\label{12-Nphotonpwfin}
\ear
Here we have introduced lightcone coordinates with
$n^\mu \equiv \frac{1}{\sqrt{2}}(0,0,1,i)$, $x^+ \equiv n\cdot x = \frac{1}{\sqrt{2}}(x^3+ix^4)$ (``light-front time''), 
and $x^- \equiv \frac{1}{\sqrt{2}}(- x^3+ix^4)$. $M$ denotes the ``effective mass'', defined by  
\bear
M^2 \equiv m^2+ \langle\langle a^2 \rangle\rangle - \langle\langle a \rangle\rangle^2 
\label{12-defM}
\ear
and 
\bear
I_\mu(\tau) \equiv \int_0^\tau   d\tau' \Bigl( a_\mu(\tau') - \langle\langle a_\mu \rangle\rangle \Bigr) 
\ear
where $\langle\langle \rangle\rangle$ stands for the worldloop average, 
$
\langle\langle f \rangle\rangle
\equiv
\frac{1}{T} \int_0^Td\tau f(\tau)\, .
$ 

Differently from the constant-field case, for the plane-wave background it seems not to be possible to reduce the transition from
scalar to spinor QED to an algebraic rule such as (\ref{rr}). The contributions from spin have to be computed by an appropriate
generalization of the Wick-contraction rule (\ref{wickferm}), namely \cite{141}
\bear
\langle \psi^\mu(\tau) \psi^\nu(\tau') \rangle = \half \mathfrak G_F^\mn(\tau,\tau')
\label{wickpsigen}
\ear
where
\begin{equation}
\hspace{-1.5em}\mathfrak G_F^\mn(\tau,\tau') 
\equiv 
\biggl\lbrace \delta^\mn + 2i n^\mu{\cal J}^\nu(\tau,\tau') + 2i {\cal J}^\mu(\tau',\tau)n^\nu
+ 2\Bigl\lbrack {\cal J}^2(\tau,\tau')-\frac{T^2}{4} \langle\langle a'\rangle\rangle^2\Bigr\rbrack  
n^\mu n^\nu\biggr\rbrace 
G_F(\tau,\tau')
\label{Gfplane}
\end{equation}
and we have further defined
\bear
J_\mu(\tau) &\equiv& \int_0^\tau d\tau' \Bigl( a'_\mu(\tau') - \langle\langle a'_\mu \rangle\rangle \Bigr) \\
{\cal J}_\mu(\tau,\tau') &\equiv& J_\mu(\tau)-J_\mu(\tau') - \frac{T}{2}\dot G (\tau,\tau')  \langle\langle a'_\mu \rangle\rangle 
\ear

\section{QED in a combined plane-wave and constant-field background}

Combining a plane-wave field with a constant one generically does not lead to a Dirac or Klein-Gordon equation that would permit
exact solutions. However, there is one well-known special case that is still solvable, defined by the existence of a Lorentz frame where
the electric and the magnetic field are parallel and the direction of propagation of the wave is in the same direction \cite{redmond,batfra,narnik,frgish-book}. 
This suggests that it should also be possible to extend the above master formulas to this case. In \cite{154} R. Shaisultanov and the author
derived this now most general known master formula for the $N$-photon amplitudes, and also elucidated why it is only in that special case that the
corresponding worldline path integrals can still be manipulated into gaussian form.

\section{Schwinger pair creation}

Going beyond the constant and plane-wave background fields usually means abandoning all hope for an exact evaluation
of the $N$-photon amplitudes or dressed propagators. Approximation schemes must be employed, of which the most universal
one in QED is the large-mass expansion of the effective action or the field-dependent propagator, corresponding to the low-energy
approximation for the corresponding photon amplitudes. 
Although the worldline formalism has been shown to be a very powerful tool for the calculation of the large-mass, or heat-kernel, expansion of the effective
action \cite{strassler2,5,gussho1,gussho2}, presently of more interest for the strong-field QED community is the application of the large-mass approximation
to the imaginary part of the effective action, since this gives the Sauter-Schwinger pair-creation rate in the field. 
In the leading approximation, the path integral is just replaced by a single trajectory, the ``worldline instanton'' \cite{afalma}. While
for a constant or plane-wave field this would, due to the gaussian character of the worldline path integrals, still be an exact procedure,
for more general fields it yields a semiclassical approximation that is expected to work well only in the large-mass limit. 

This application of the worldline formalism to Schwinger pair creation is already a well-developed subject that has spawned a
considerable literature \cite{63,64,dunwan,dugiscPRL,ilderton,schsch-schwinger,akamoo,degtor2}, 
and would really require a separate talk. Thus let me just shortly summarize its most salient features:

\benn

\item
It localizes the path integral on a solution of the classical equation of motion (the Euclidean Lorentz force equation). 

\item
It usually works well only for the imaginary part of the effective action, not for the real one. 

\item
It is equivalent to WKB but technically superior in a number of aspects.  

\item
For planar fields, there is a universal analytic formula for the pair creation rate involving a single master integral. If one can calculate this integral, It is {\it not necessary to
compute the worldline instanton}. This includes the fluctuation prefactor. 

\item
For non-planar fields, usually one will have to go for a numerical determination of the worldline instanton(s) \cite{dunwan}. 

\enn

Besides Sauter-Schwinger pair-production, the semiclassical approximation to the worldline path integral has also been applied to other QED processes
such as Breit-Wheeler pair production \cite{satunin,degtor1}.

\section{Outlook}

The worldline formalism at this stage is ready for applications to

\begin{itemize}

\item
Schwinger pair creation in arbitrary electric fields. 

\item
Closed-loop and open-line processes in arbitrary constant fields. 

\item
Closed-loop processes in plane-wave backgrounds. 

\end{itemize}

In particular, it applies to (linear and non-linear) Compton scattering and to light-by-light scattering in either a constant or plane-wave field.

\end{document}